# A sterile neutrino search at compact materials irradiation facility

Liangwen Chen[1], Han-Jie Cai[1], Emilio Ciuffoli[1], Jarah Evslin[1,3], Fen Fu[1,2], Sheng Zhang[1], Xurong Chen[1,2], Lei Yang[1,2,a], Wenlong Zhan[1]

[1] Institute of Modern Physics, CAS, Lanzhou 730000, China
[2] School of Nuclear Science and Technology, University of the Chinese Academy of Sciences, Beijing 100049, China
[3] University of the Chinese Academy of Sciences, Beijing 100049, China

**Abstract** The compact material irradiation facility (CMIF) is a current project in China that will provide a compact deuteron-beryllium neutron source. The target of this facility will be an intense and compact Isotope Decay-At-Rest (IsoDAR) neutrino source. In this paper, we propose to test the sterile neutrino hypothesis using CMIF as the neutrino source. At CMIF platform, the electron antineutrino production rate can be up to $2.0 \times 10^{19}$ per day. When paired with an 80 t liquid scintillator detector to study short baseline electron antineutrino disappearance, the inverse beta decay (IBD) event rate is large enough to investigate the parameter ranges of interest for neutrino anomalies. Our sensitivity analysis shows that a short baseline experiment at this platform will provide a very competitive sterile neutrino search, especially in the high-$\Delta m^2$ region ($\Delta m^2 > 10\,\text{eV}^2$).

## 1 Introduction

Neutrino oscillations have been observed at solar and atmospheric neutrino experiments [1,2], and confirmed independently by accelerator and reactor neutrino experiments [3–9]. Since then, the three mixing angles and the two mass squared differences have been determined with good precision and the three-flavour neutrino oscillation paradigm has been well established. However, some anomalies have been observed in neutrino experiments with different sources, detector principles and detection channels [10–18] and may ultimately challenge the standard three-flavor neutrino paradigm. These anomalous results may be a sign of new physics beyond Standard Model, with the most popular extension being the addition of sterile neutrino states with masses at the eV scale.

During the past few years, various neutrino experiments with different sources and detection concepts have been running to directly explore the favored parameter space for the sterile neutrino and the relevant parameter space is restricted by several null results [19–25]. Recently, evidence for one of these anomalies, the reactor anomaly, has been weakened by the Daya Bay's reactor fuel cycle measurements [26] and also by the observed shoulder at 5 MeV in the reactor neutrino spectrum [27–29], which is in contradiction with the theoretical reactor flux prediction of Ref. [30]. In spite of this, the sterile neutrino hypothesis is only mildly disfavored and understanding the reactor anomaly remains a challenge [31,32].

In addition to the intrinsic interest in determining the existence of new phenomena and particles, sterile neutrino states are a key issue in searches for new physics beyond the Standard Model [33,34], and their existence and properties would also have important implications in astrophysics [35] and cosmology [36,37]. In recent years, short baseline neutrino experiments adopting various neutrino sources have been proposed for sterile neutrino searches [38–46]. Among them, the Isotope Decay-At-Rest (IsoDAR) proposal [38] has aroused attention. They propose to use a 60 MeV, 10 mA proton beam to produce neutrons in a beryllium target. These neutrons travel into a cylindrical sleeve of isotopically pure, solid $^7$Li which is 1.5 m long and 2 m in outer diameter. The neutrons are absorbed by the $^7$Li, yielding $^8$Li whose $\beta$ decay provides electron antineutrinos for the experiment. To obtain a sufficient capture efficiency, the isotopic purity of the $^7$Li converter should be quite high, at least 99.99%; this would increase the cost of the experiment. A volume of graphite is used to act as a neutron reflector outside the $^7$Li converter. Lithium hydroxides (LiOD, LiOD · $D_2O$), lithium deuteride (LiD) and FLiBe molten salt [47–50] have been also proposed in place of metallic lithium. It should be noticed that these IsoDAR proposals demand a complicated target configuration and years of dedicated beam time.

The compact materials irradiation facility (CMIF) project will be a high-energy, high-flux neutron source for material

[a] e-mail: lyang_imp@outlook.com

Springer

irradiation research, which is indispensable for the long-term Accelerator Driven System (ADS) project in China [51–55]. In this project, a 10 mA deuteron beam with an energy of 50 MeV will bombard a beryllium target. In addition to the neutron flux, a considerable amount of $^8$Li can be produced directly in the target as a by-product of the collisions, without the need for a $^7$Li converter. Evidence for the production of this radioactive nucleus in the bombardment of beryllium by a low energy deuteron beam was obtained already in 1954 [56]. For the reaction $^9$Be(d, x)$^8$Li, the calculated Q-value is $-11.37 \pm 0.07$ MeV, corresponding to a threshold deuteron energy of 13.9 MeV. Thus, the production of $^8$Li at CMIF is feasible from the perspective of beam energy. It should be noted that in this proposal a complicated target configuration for $^8$Li production is not required, reducing the cost of the experiment. Since no additional modification is demanded for the accelerator-target configurations, the proposed experiment can start operation during the commissioning of the first-stage of CMIF.

In this paper, we will study the feasibility of a sterile neutrino search at CMIF. The paper is organized as follows. First, the flux intensity of the electron antineutrino at CMIF will be evaluated in Sect. 2. The experimental setup will be described in Sect. 3. Then the event rate distribution in the detector is found in Sect. 4 and the sensitivity of a sterile neutrino search at this facility will be investigated in Sect. 5. Finally a short summary will be given in Sect. 6.

## 2 Flux intensity of electron antineutrinos at CMIF

For the estimation of the flux intensity of the electron antineutrino, the simulation of the beam-target interaction is performed using the GEANT4.10 toolkit [57]. In the GEANT4 simulation, the INCL++ model [58,59] which can give a comparatively better description of the neutron production in the deuteron-beryllium reaction [52] is adopted. Owing to the fact that the three radioactive nuclei are mainly produced as residual nuclei after the emission of neutrons and protons, to a certain extent, a comparatively high reliability can be anticipated for the estimation of isotope production if the neutron yield prediction is accurate enough.

Many isotopes that could emit electron antineutrinos by beta decay can be produced in the target. However, in order to test the hypothesis of sterile neutrino states with masses at the eV scale, considering that the baseline can be on the order of several meters and that a veto of 6 MeV is adopted for the neutrino energy (as will be stated in the following discussions), only neutrinos with energies of several MeV will be of interest. Nearly all of these are produced in decays of $^6$He, $^8$Li and $^9$Li. Therefore, the contributions from other isotopes are neglected in the analysis. As the Be target of CMIF is quite thin ($\sim$ 2 g/cm$^2$), the production of these three

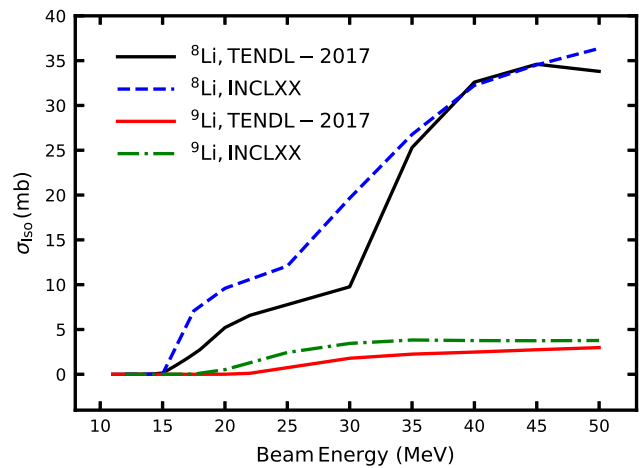

**Fig. 1** Comparsion between the INCL++ simulation and the TENDL-2017 Nuclear data library [60] for the production cross section of $^8$Li and $^9$Li in low energy deuteron-beryllium reactions

radioactive isotopes is mainly due to the primary reaction between the deuteron beam and the target. Figure 1 compares the INCL++ estimation and the TENDL-2017 Nuclear data library for the production cross section of $^8$Li and $^9$Li in D-Be reactions below 50 MeV. A good consistency in the $^8$Li cross section is seen above 35 MeV. Note that with the beam-target configuration of CMIF, more than 70% of $^8$Li are produced by reactions induced by deuterons above 35 MeV, which means that the reliability above 35 MeV is most important. As for $^9$Li, the INCL++ estimation tends to be larger than TENDL-2017 [60]. This is acceptable because $^9$Li only contributes $\sim$ 10% of electron antineutrinos, and so this higher behavior (by tens of percent) will have only a modest effect on the estimation of the total flux.

The flux intensities of the electron antineutrinos from different isotopes and the total are shown in Fig. 2. The spectrum of the electron antineutrinos from $^6$He has a peak at 2 MeV while that from $^8$Li and $^9$Li have a similar energy range and their shapes are quite similar. According to the INCL++ simulation, the integrated flux intensity will be $2.0 \times 10^{19}$ per day above the 1.8 MeV inverse beta decay (IBD) threshold. For the mitigation of the backgrounds, which will be described in detail later, we consider a low energy cut at 6 MeV for neutrinos. Therefore, the neutrinos contributed by $^6$He are irrelevant. In the neutrino energy range of 6–14 MeV, the integrated flux intensity will be $8.2 \times 10^{18}$ per day.

The production rate of $^8$Li and $^9$Li will be measured experimentally when the sterile neutrino search at CMIF is to proceed. Both of these isotopes are easily identified [56,61,62] as a reaction product because of their characteristic beta decay followed by the breakdown of residual $^8$Be nucleus into two alpha particles (For $^9$Li, there is a branching ratio of 50.5% for the decay to $^9$Be.). In the following analysis, a flux uncertainty of 20% will be adopted.

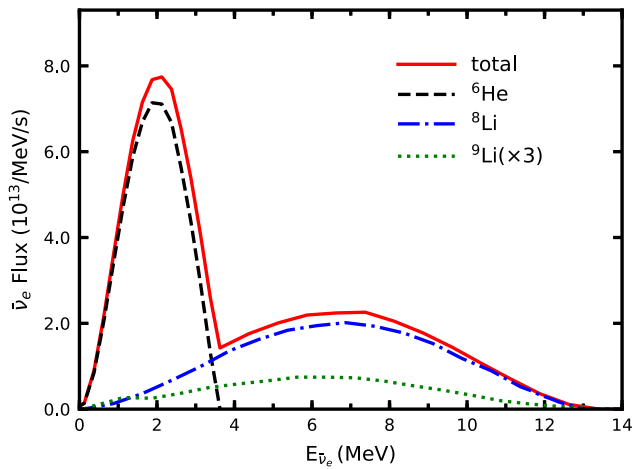

**Fig. 2** The flux intensity of electron antineutrinos from the isotopes $^6$He, $^8$Li and $^9$Li. Their sum is the red solid curve. For visual comparison, the $^9$Li contributions are scaled by a factor of 3

## 3 Proposed experimental setup

In this paper, we will consider a test of the $3 + 1$ sterile neutrinos hypothesis. The electron antineutrino disappearance probability in this scenario can be parametrized as

$$P_{ee}(L, E) = 1 - \sin^2(2\theta) \sin^2 \left( 1.27 \frac{\Delta m^2 [\text{eV}^2] L[\text{m}]}{E[\text{MeV}]} \right), \quad (1)$$

where $\Delta m^2$ is the relevant squared mass splitting, $\theta$ is the effective mixing angle, $L$ is the neutrino travel distance, and $E$ is the neutrino energy.

We propose to pair the neutrino source with liquid scintillator detectors, which should have a total fiducial mass of about 80 t. To simplify the analysis, we consider four of the Daya Bay detectors. The same muon veto systems as Daya Bay are also adopted. Each of the Daya Bay detectors is about 5 m in diameter and 4 m in height. The fiducial volume of each detector is 3 m in diameter and 3 m in height, and the fiducial mass is about 20 t and target proton number is about $1.4 \times 10^{30}$ [63]. The energy resolution and the position resolution for the Daya Bay detectors are taken as $9\%/\sqrt{E(\text{MeV})}$ and 15 cm, respectively [40].

The most precise experiments that will study sterile neutrinos in the disappearance channel will employ reactor neutrinos [45,64,65]; whose energy is significantly lower than IsoDAR neutrinos. The oscillation length for a neutrino of energy $E$ is

$$L = 4\pi \frac{E}{\Delta M^2} \quad (2)$$

where $\Delta M^2$ is the squared mass splitting between the neutrino mass eigenstates and we have adopted natural units

$\hbar = c = 1$. Therefore if the neutrino energy, which is a feature of the experiment, and the mass squared splitting, which depends on the model, are increased by the same factor, then the oscillation length will be unchanged. If the oscillation length is of order the size of the source or the position resolution of the detector, then the oscillations will be suppressed [66]. In our case, the average neutrino energy is two to three times greater than that of reactor neutrinos, while the source size is comparable. Therefore, with the same position resolution as another experiment, we are sensitive to neutrino oscillations in a model with a $\Delta M^2$ which is two to three times greater than that in a reactor experiment with a similar event rate and background suppression. As the neutrino source is compact, and considering the constraints from the detector geometry and the shielding blocks, we suppose a baseline, which is the distance between the detector center to the neutrino source, of 4 m.

The overburden of the CMIF project is supposed to be in the range of 5–30 m. The best choice of overburden will be discussed further in the following analysis. At an energy of several MeV, the $\bar{\nu}_e$ events are detected via the IBD reaction, where a positron and a neutron is produced. The positron will promptly annihilate with an electron, and in a Daya Bay detector the neutron will be thermalized and captured by Gd nuclei followed by a gamma cascade with total energy of about 8 MeV. The time difference between the two signals is about $30\,\mu s$. Given the IBD signal, the following three cosmogenic muon induced backgrounds will be inevitable: the accidental background, the fast neutron background and the $^9$Li background. There are 14 nuclear reactors which are either operational or under construction in Guangdong province, where the CMIF project will be located. Among them, the nearest will be the Taiping Ling reactor cluster. This will eventually consist of six reactors, all at a distance of about 5 km from the CMIF project. Its construction was confirmed in early 2019 and it will be in operation in 2025. As a result, reactor antineutrinos will contribute to the backgrounds. As reported in Ref. [67], the muon induced backgrounds can be considerably mitigated by a low energy veto. In the context of this paper, the neutrino events with true energy in the range of 6–14 MeV are selected. Within this range, the accidental background can be omitted. Therefore, three kinds of backgrounds will be included in our analysis: the reactor neutrino background, the $^9$Li induced background, and fast neutron induced background.

## 4 Event rate distributions

To investigate the physical potential of a sterile neutrino search at this facility, we need to first obtain the event rate. In this section, we will calculate the IBD event rate and the background event rate distributions mentioned above.

In the Daya Bay liquid scintillator detectors, electron antineutrinos are detected via the IBD process $\bar{\nu}_e + p \to e^+ + n$. The IBD cross section can be parameterized as [68]

$$\sigma^{\text{IBD}}(E) = p_e E_e E^{-0.07056 + 0.02018 \ln E - 0.001953 \ln^3 E} \times 10^{-43} \text{ (cm}^2\text{)},$$

$$E_e = E + m_p - m_n \approx E - 1.293 \text{ MeV}, \quad (3)$$

where $p_e$ and $E_e$ are the momentum and total energy of the positron. The neutrino energy is denoted as $E$. $m_p$ and $m_n$ are the proton and neutron masses, respectively. All of these energies and masses are expressed in MeV. In IBD process, the neutrino energy threshold is given by

$$E_{\text{thre}} = \frac{(m_n + m_e)^2 - m_p^2}{2 m_p} \sim 1.806 \text{ MeV}. \quad (4)$$

Since the neutrino source is much smaller than the detector, it is taken to be a point source in the following analysis. The expected number of IBD events in the energy bin $j$ and distance (the distance between the neutrino source and the detection point) bin $i$ in each detector can be written as

$$S^{ij} = \int dV_i \frac{n_p}{V} \int dE_j \frac{\Phi_{\bar{\nu}_e}(E_j) P_{ee}(L_i, E_j) \sigma^{\text{IBD}}(E_j)}{4 \pi L_i^2}, \quad (5)$$

where $n_p$ is the total number of target protons in the detector, $V$ the total volume of the detector. $\Phi_{\bar{\nu}_e}$ is the $\bar{\nu}_e$ flux from the source. In the energy bin $j$, the neutrino energy is denoted by $E_j$. In the distance bin $i$, the volume is denoted by $V_i$ and the travel distance of the corresponding neutrino is denoted by $L_i$. As mentioned above, the Daya Bay detectors are adopted and the energy and position resolution parameters are taken to be $9\%/\sqrt{E(\text{MeV})}$ and 15 cm. According to these conditions, the energy bin width and distance bin width are chosen to be 0.2 MeV and 0.25 m, respectively, which are comparable with the corresponding resolution parameters. $P_{ee}(L_i, E_j)$ is the neutrino oscillation probability, as described in Eq. (1). $S^{ij}$ will be subjected to Gaussian energy and position resolution smearing.

In Fig. 3, the energy dependence of signal event rates is shown in the upper panel and the position dependence in the lower panel. $L$ denotes the neutrino travel distance. In the upper panel, the bump at $\sim 3$ MeV is mainly due to the electron antineutrino contribution from the $^6$He decay, which will not be considered in the following sensitivity analysis due to the energy cut adopted. In the neutrino energy range of 6–14 MeV, the total signal event rates in each Daya Bay detector with a baseline of 4 m are 23.6 per day. With a baseline of 4 m, the event rates are mainly in the distance range of 2.5–5.5 m.

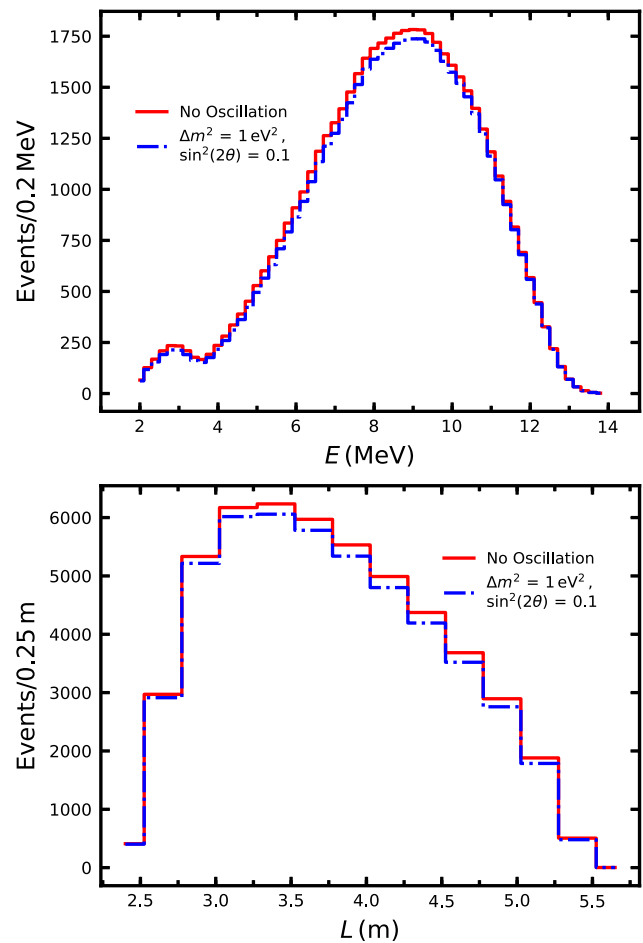

**Fig. 3** The energy and position dependence of the event rates in a Daya Bay detector with a baseline of 4 m. The red line is the IBD event rate with no oscillation, and the blue dashed curve is the IBD event rate with $\Delta m^2 = 1 \text{ eV}^2$ and $\sin^2(2\theta) = 0.1$. A five year run and duty factor of 90% are adopted

As mentioned above, the reactor neutrino background and muon induced backgrounds are relevant. The total rates for the reactor antineutrino background are estimated by assuming the same average thermal power as at the Daya Bay reactor experiment [63]. The measured energy spectrum of the reactor antineutrino is known well and is taken from Refs. [69,70]. The total rates of the $^9$Li and fast neutron induced backgrounds are taken from Ref. [71]. The spectrum of the $^9$Li induced background, which are approximately universal for different overburdens, is adopted from Refs. [63,69]. Unlike the $^9$Li and reactor neutrino backgrounds, the measured spectrum of the fast neutron induced background varies appreciably with the overburden, as was investigated in Ref. [71]. It should be noticed that in Ref. [71], the energy spectrum of fast neutron induced background is only shown within the neutrino energy range of [6 MeV, 14 MeV].

Above 6 MeV, the total rate of the reactor antineutrino background is about 1.0 events per day in each Daya Bay

**Table 1** The total rates per day for fast neutron induced and $^9$Li induced backgrounds for each Daya Bay detector with different overburdens [71]. Here the event rates are obtained with the neutrino energy in the range of 6–14 MeV

| Overburden (m) | Fast neutron | $^9$Li |
|---|---|---|
| 5 | 189.0 | 21.9 |
| 10 | 117.8 | 12.5 |
| 20 | 83.2 | 8.3 |
| 30 | 49.0 | 4.2 |

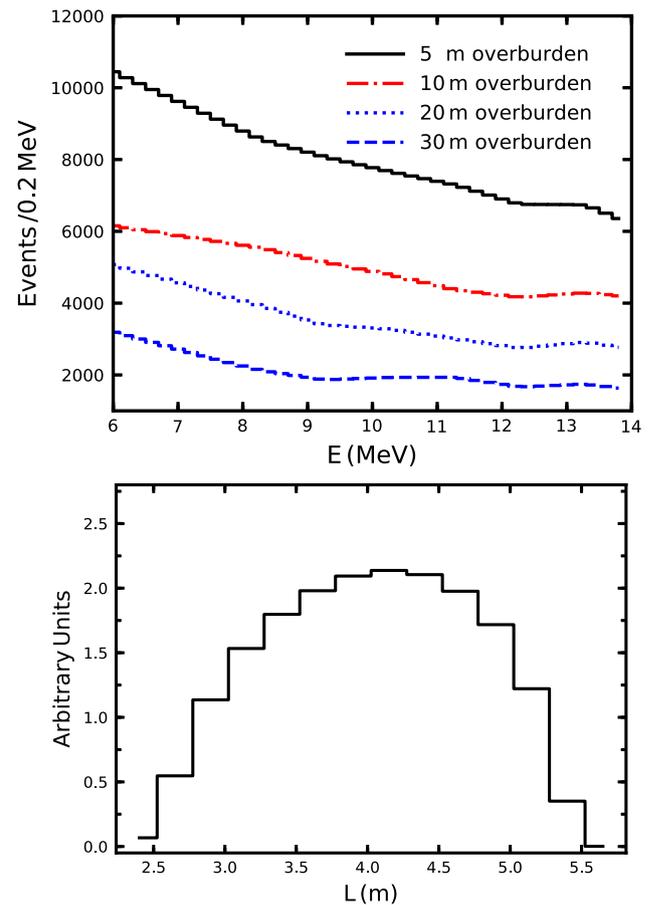

**Fig. 4** The upper panel shows the energy spectra for fast neutron induced backgrounds with different overburdens. The normalized energy spectra are adopted as in Ref. [71]. The neutrino energy is shown within the energy range of [6 MeV, 14 MeV]. The lower panel shows the distance dependence of background event rates, which are universal for all the backgrounds

detector. The total rates of the fast neutron induced and $^9$Li induced backgrounds in each Daya Bay detector with different overburdens are shown in Table 1. Since the signal rate is 23.6 per day, as mentioned above, it can be concluded that the fast neutron induced background is the most relevant, which is about eight times the signal rate with overburden of 5 m and $\sim$ 4 times the signal rate with overburden of 20 m. The $^9$Li induced background total rate is about 0.1 that of the fast neutron induced background. Compared to the other two backgrounds, the reactor antineutrino background is quite small in the analysed overburden ranges. In the upper panel of Fig. 4, the energy spectra of fast neutron backgrounds with different overburdens are shown [71]. It can be seen that the energy spectrum is quite uniform in the plotted range compared with those of the signal and the $^9$Li induced background. The energy spectrum of the $^9$Li is adopted from Fig. 3.11 in Ref. [69] and is not shown here. It has a peak around a neutrino energy of about 7 MeV, decreases quickly and has an endpoint at a neutrino energy of about 13 MeV. Since the background events are distributed uniformly in the detectors, different backgrounds have the same distance dependence. In the lower panel of Fig. 4, the distance dependence of the background events is shown. Since the baseline is chosen to be 4 m, as mentioned above, due to the finite detector size, some neutrinos travel less than 4 m and so the distance dependence curve has a peak around 4 m. It should be considered that, due to the significant geometrical effect on the neutrinos from CMIF, the distance distribution in Fig. 2 for the signal has a peak beneath that of the backgrounds.

## 5 Sensitivity of sterile neutrino search

In our statistical analysis, a five year running time and 90% duty factor are adopted. We assume a flux normalization uncertainty $\sigma_s$ of 20% for the neutrino source, as discussed above, normalization uncertainties $\sigma_r$ of 1%, $\sigma_l$ of 10%, $\sigma_f$ of 10% for the reactor background from the reactor complex, the $^9$Li induced background, and the fast neutron induced background, respectively. The resulting $\chi^2$ is defined as the following function:

$$\chi^2 = \sum_{\mathrm{AD}} \sum_{i,j} \frac{\left(N_{\mathrm{obs}}^{\mathrm{AD},ij} - N_{\mathrm{exp}}^{\mathrm{AD},ij}\right)^2}{N_{\mathrm{obs}}^{\mathrm{AD},ij}} + \left(\frac{\alpha_s}{\sigma_s}\right)^2 + \left(\frac{\alpha_r}{\sigma_r}\right)^2 + \left(\frac{\alpha_l}{\sigma_l}\right)^2 + \left(\frac{\alpha_f}{\sigma_f}\right)^2, \qquad (6)$$

where AD sums over the detectors, the index $i$ labels the neutrino energy bins, the index $j$ labels the distance bins, $\alpha_s$, $\alpha_r$, $\alpha_l$, $\alpha_f$ are the nuisance parameters accounting for the normalization uncertainties of the neutrino flux, the reactor $\bar{\nu}_e$ background, the $^9$Li induced background and the fast neutron induced background, respectively. $N_{\mathrm{obs}}^{\mathrm{AD},ij}$ is the observed number of events in each bin assuming no sterile neutrino oscillation. $N_{\mathrm{exp}}^{\mathrm{AD},ij}$ is the expected number of events in each bin including possible sterile neutrino oscillation. It is the sum of the events from the neutrino source ($S_{\mathrm{exp}}^{\mathrm{AD},ij}$), the background from the reactor neutrinos ($R^{\mathrm{AD},ij}$), the $^9$Li induced background ($L^{\mathrm{AD},ij}$) and fast neutron induced back-

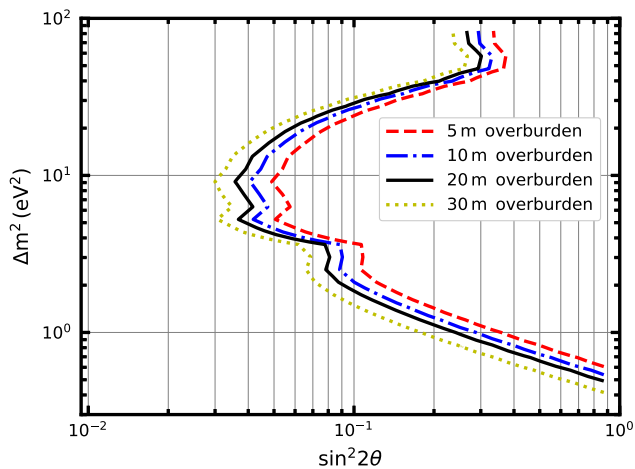

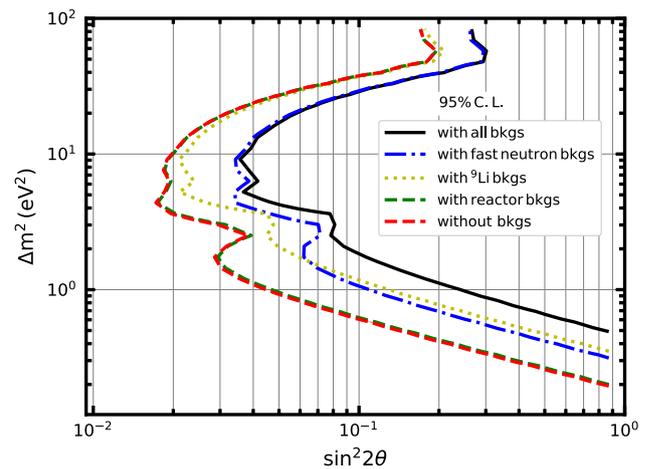

**Fig. 5** The 95% C.L. sensitivity curves of the sterile neutrino search with different overburdens. The red dashed curve corresponds to an overburden of 5 m, the blue dash-dotted curve to an overburden of 10 m, the black solid curve to 20 m, and the yellow dotted curve to 30 m. A five year run and 90% duty factor are assumed

**Fig. 6** The 95% C.L. sensitivity curves of the sterile neutrino search with different backgrounds included. The overburden is fixed to 20 m. The black solid curve corresponds to the analysis with all three backgrounds considered. The blue dash-dotted curve considers only the fast neutron backgrounds, while the yellow dotted curve considers only $^9$Li backgrounds and the green dashed curve only reactor backgrounds. The red dashed curve corresponds to the no background analysis

ground ($F^{AD,ij}$), that is,

$$N_{\exp}^{AD,ij} = (1+\alpha_s)S_{\exp}^{AD,ij} + (1+\alpha_r)R^{AD,ij} \\ + (1+\alpha_l)L^{AD,ij} + (1+\alpha_f)F^{AD,ij}. \quad (7)$$

The oscillation probability is parameterized as in Eq. (1). We scan the parameter space ($\sin 2\theta$, $\Delta m^2$); for each point the $\chi^2$ function (defined as in Eq. (5)) is minimized over nuisance parameters by using the Powell method. The resulting $\chi^2$ function gives the confidence level at which the non-oscillation hypothesis can be ruled out.

As mentioned above, the overburden of the CMIF project is supposed to be in the range of 5–30 m. Here, we investigate the effect of overburden on the sensitivity and choose the best overburden for the sterile neutrino search. In Fig. 5, the sensitivity curves for a sterile neutrino search with different overburdens are shown. As expected, when the overburden increases, the backgrounds decrease and the exclusion regions become larger.

In Fig. 6, the effects of different backgrounds in the sterile neutrino search are shown, assuming a 20 m overburden. It can be seen that when $\Delta m^2$ is larger than 10 eV$^2$, the most relevant background is the fast neutron induced background. This is because the sensitivity in this region is mainly controlled by the events with larger neutrino energy, and the fast neutron induced background is distributed more uniformly than the other two backgrounds and dominates when neutrino energy is large. When $\Delta m^2$ is lower than several eV$^2$, the $^9$Li induced and fast neutron induced backgrounds both play important roles. Since in this area the sensitivity is controlled mainly by the events with lower neutrino energy, the $^9$Li induced background, whose spectrum has a peak around 7 MeV and decreases sharply, and fast neutron induced background both play important roles. The effect of the reactor antineutrino background can be neglected in the plotted regions, as is expected, because the total rates are quite tiny and the spectra are peaked at neutrino energy of about 4 MeV, which is below the neutrino energy cut in the analysis, and decreases sharply as the energy approaches ∼ 8 MeV.

The 95% exclusion region in ($\Delta m^2$, $\sin^2 2\theta$) parameter space at CMIF is presented in Fig. 7, assuming again a 20 m overburden. The 95% C.L. allowed region from the combination of reactor neutrino experiments, Gallex and Sage calibration sources experiments, and the MiniBoone [16], as well as the expected 95% C.L. exclusion curves of both the DANSS 1 year running and SoLiD phase II 3 years running [64,65,72] are shown, too. Compared with the other shown oscillation experiments, one sees that CMIF provides the best bounds on sterile neutrino mixing at mass squared differences above about 10 eV$^2$, covering the area of interest for anomalies over a sizeable range of mass splittings. At high $\Delta m^2$ the sensitivity of CMIF exceeds these other oscillation experiments because, as was mentioned before, they use neutrinos from nuclear power plants, which have lower energy than IsoDAR neutrinos and so have shorter oscillation length at high mass splitting. The sensitivity in this region is one of the main advantages of this proposed sterile neutrino search at CMIF, making it complementary to other proposed oscillation experiments for sterile neutrino searches. In addition to oscillation experiments, experiments designed to measure absolute neutrino masses are also expected to be sensitive to sterile neutrinos [73–79], for example, the KATRIN experiment [74,75,79]. The expected 95% C.L. sensitivity of the KATRIN experiment [79] is also shown in Fig. 7. One sees that the sensitivity of KATRIN experiment is slightly tighter than our sensitivity to sterile neutrinos near 10 eV$^2$. However, as an oscillation experiment our systematics will be quite different. For example, the sensitivity of absolute mass experiments to sterile neutrinos depends on the absolute neutrino masses [78]. Needless to say, multiple channels will be required for a convincing discovery of sterile neutrinos.

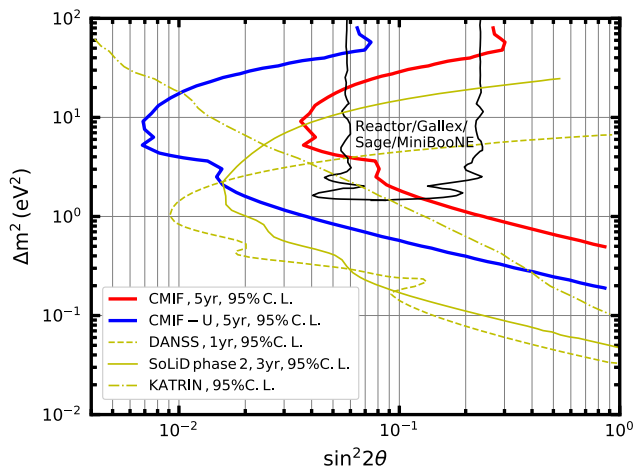

**Fig. 7** The exclusion limits that can be obtained with CMIF and CMIF-U at 95% C.L.. The area inside of the black curve is the allowed region at 95% C.L. in the parameter space from the combination of reactor neutrino experiments, Gallex and Sage calibration sources experiments and MiniBooNE [16]. The 95% C.L. exclusion curves of the DANSS, SoLiD and the KATRIN [64,65,72,79] are also presented as the yellow lines

In fact, the total $\bar{\nu}_e$ flux intensity in this sterile neutrino search proposal is quite moderate compared to those of other congenerous proposals. However, it can be enhanced by a factor of more than ten after the upgrade of CMIF [51,52] (which we will refer to as CMIF-U), which will be carried out in the second phase of the neutron source project and will have a beam energy of 250 MeV [51,52]. Here a flux of $8 \times 10^{19}$ per day, which will be easily obtained at CMIF-U, is supposed and the corresponding sensitivity curve is also given in Fig. 7. In this case, the exclusion limits of CMIF-U are much tighter than those of CMIF, and are tighter than those of competing experiments at a very wide range of mass splittings.

## 6 Conclusion

In this paper, we proposed to test the sterile neutrinos hypothesis at the CMIF platform, without any additional modifications to the original accelerator-target configuration. Our proposal uses IsoDAR neutrinos, which are higher energy than reactor neutrinos, allowing it to probe higher mass splittings than reactor neutrino experiments. IsoDAR proposals are generally expensive as they require tons of highly isotopically pure $^7$Li and years of dedicated beam time. Ours requires neither, the experiment will use neutrinos that will anyway be produced during the planned operation of CMIF.

With the beam-target parameters at CMIF, the $\bar{\nu}_e$ source is compact and the $\bar{\nu}_e$ production rate can be up to $2.0 \times 10^{19}$ per day. When paired with four Daya Bay detectors, the overburden should be at least 20 m in order to obtain a competitive sterile neutrino search ability. We have considered this overburden and a running time of 5 years and duty factor of 90%. If there is a single sterile neutrino with a mass splitting between 5 eV$^2$ and 20 eV$^2$, then the exclusion limits of this proposal cover the entire allowed region of sterile neutrino mixing parameters from the combination of reactor neutrino experiments, Gallex and Sage calibration sources experiments, and the MiniBoone [16]. It is complimentary to other oscillation experiments for sterile neutrino searches as it is most sensitive at mass splittings of $\Delta m^2 \sim 10$ eV$^2$, and it provides another channel for sterile neutrino searches in the high mass region in addition to $\beta$ decay experiments like KATRIN [74] and electron capture experiments like ECHo [75]. It will contribute a convincing test of the sterile neutrino hypothesis. After the upgrade CMIF-U, our proposal covers the entire region of interest at 95% confidence up to a mass splitting of 50 eV$^2$.

The CMIF project in China may provide great opportunities for a sterile neutrino search. In particular, CMIF-U can provide a decisive test of disappearance channel neutrino anomalies.

**Acknowledgements** We thank Yu-Feng Li, Jian Tang and Jiajie Ling for helpful discussions. This work is supported in part by the Strategic Priority Research Program of the Chinese Academy of Sciences, Grant no. XDA03030100. Liangwen Chen is supported by the National Postdoctoral Program for Innovative Talents (BX201700258). Han-Jie Cai is supported by NSFC Grant no. 11805253. Sheng Zhang is supported by NSFC no. 11705256. JE is supported by the CAS Key Research Program of Frontier Sciences Grant QYZDY-SSW-SLH006 and the NSFC MianShang Grants 11875296 and 11675223. EC is supported by NSFC Grant no. 11605247, and by the Chinese Academy of Sciences Presidents International Fellowship Initiative Grant no. 2015PM063. JE and EC thank the Recruitment Program of High-end Foreign Experts for support.

**Data Availability Statement** This manuscript has no associated data or the data will not be deposited. [Authors' comment: The data that support the findings of this study are available from the corresponding author, [Yang Lei], upon reasonable request.]

Funded by SCOAP³.